\documentclass[12pt,hyper,notoc]{JHEP3}
\usepackage{graphics}
\usepackage{psfrag}
\usepackage{amsmath}
\usepackage{amsthm}
\usepackage{amssymb}
\usepackage{epsfig}
\usepackage{euscript}
\usepackage{array}
\usepackage{cite}
\usepackage{cancel}
\usepackage{empheq}
\usepackage{verbatim}

\theoremstyle{definition}

\theoremstyle{remark}

\newcounter{multieqs}

\newcommand{\be}{\begin{equation}}
\newcommand{\ee}{\end{equation}}
\newcommand{\eq}[1]{(\ref{#1})}
\newcommand{\bit}{\begin{itemize}}  \newcommand{\eit}{\end{itemize}}

\newcommand{\bm}[1]{\mbox{\boldmath $#1$}}
\newcommand{\rf}[1]{(\ref{#1})}

\def\bd{\begin{document}}
\def\ed{\end{document}}
\def\nn{\nonumber}
\def\bea{\begin{eqnarray}}
\def\eea{\end{eqnarray}}
\let\bm=\bibitem

\def\la{\langle}
\def\ra{\rangle}

\def\npb#1#2#3{Nucl. Phys. {\bf{B#1}} #3 (#2)}
\def\plb#1#2#3{Phys. Lett. {\bf{#1B}} #3 (#2)}
\def\prl#1#2#3{Phys. Rev. Lett. {\bf{#1}} #3 (#2)}
\def\prd#1#2#3{Phys. Rev. {D \bf{#1}} #3 (#2)}
\def\cmp#1#2#3{Comm. Math. Phys. {\bf{#1}} #3 (#2)}
\def\cqg#1#2#3{Class. Quantum Grav. {\bf{#1}} #3 (#2)}
\def\nppsa#1#2#3{Nucl. Phys. B (Proc. Suppl.) {\bf{#1A}}#3 (#2)}
\def\ap#1#2#3{Ann. of Phys. {\bf{#1}} #3 (#2)}
\def\ijmp#1#2#3{Int. J. Mod. Phys. {\bf{A#1}} #3 (#2)}
\def\rmp#1#2#3{Rev. Mod. Phys. {\bf{#1}} #3 (#2)}
\def\mpla#1#2#3{Mod. Phys. Lett. {\bf A#1} #3 (#2)}
\def\jhep#1#2#3{J. High Energy Phys. {\bf #1} #3 (#2)}
\def\atmp#1#2#3{Adv. Theor. Math. Phys. {\bf #1} #3 (#2)}

\def\N{{\cal N}}
\def\sst{\scriptscriptstyle}
\def\thetabar{\bar\theta}
\def\Tr{{\rm Tr}}
\def\one{\mbox{1 \kern-.59em {\rm l}}}

\def\a{\alpha}      \def\da{{\dot\alpha}}  \def\dA{{\dot A}}
\def\b{\beta}       \def\db{{\dot\beta}}  
\def\g{\gamma}  \def\G{\Gamma}  \def\dc{{\dot\gamma}}  
\def\d{\delta}  \def\D{\Delta}  \def\ddt{\dot\delta}  
\def\e{\epsilon}        \def\ve{\varepsilon}  
\def\f{\phi}    \def\F{\Phi}    \def\vvf{\f}  
\def\h{\eta}  
\def\k{\kappa}  
\def\l{\lambda} \def\L{\Lambda}  
\def\m{\mu} \def\n{\nu}  
\def\o{\omega}  
\def\p{\pi} \def\P{\Pi}  
\def\r{\rho}  
\def\s{\sigma}  \def\S{\Sigma}  
\def\t{\tau}  
\def\th{\theta} \def\Th{\Theta} \def\vth{\vartheta}  
\def\X{\Xeta}  
\def\z{\zeta}  

\def\na{\nabla}  
\def\cA{{\cal A}} \def\cB{{\cal B}} \def\cC{{\cal C}}  
\def\cD{{\cal D}} \def\cE{{\cal E}} \def\cF{{\cal F}}  
\def\cG{{\cal G}} \def\cH{{\cal H}} \def\cI{{\cal I}}  
\def\cJ{{\cal J}} \def\cK{{\cal K}} \def\cL{{\cal L}}  
\def\cM{{\cal M}} \def\cN{{\cal N}} \def\cO{{\cal O}}  
\def\cP{{\cal P}} \def\cQ{{\cal Q}} \def\cR{{\cal R}}  
\def\cS{{\cal S}} \def\cT{{\cal T}} \def\cU{{\cal U}}  
\def\cV{{\cal V}} \def\cW{{\cal W}} \def\cX{{\cal X}}  
\def\cY{{\cal Y}} \def\cZ{{\cal Z}}

\def\ua{\underline{\alpha}}  
\def\uc{\underline{\phantom{\alpha}}\!\!\!\gamma}  
\def\um{\underline{\mu}}  
\def\ud{\underline\delta}  
\def\ue{\underline\epsilon}  
\def\una{\underline a}\def\unA{\underline A}  
\def\unb{\underline b}\def\unB{\underline B}  
\def\unc{\underline c}\def\unC{\underline C}  
\def\und{\underline d}\def\unD{\underline D}  
\def\une{\underline e}\def\unE{\underline E}  
\def\unf{\underline{\phantom{e}}\!\!\!\! f}\def\unF{\underline F}  
\def\unm{\underline m}\def\unM{\underline M}  
\def\unn{\underline n}\def\unN{\underline N}  
\def\unp{\underline{\phantom{a}}\!\!\! p}\def\unP{\underline P}  
\def\unq{\underline{\phantom{a}}\!\!\! q}  
\def\unQ{\underline{\phantom{A}}\!\!\!\! Q}  
\def\unH{\underline{H}}  
  
\def\As {{A \hspace{-6.4pt} \slash}\;}  
\def\bs {{b \hspace{-6.4pt} \slash}\;}  
\def\Ds {{D \hspace{-6.4pt} \slash}\;}
\def\Gts {{\Gt \hspace{-6.4pt} \slash}\;}
\def\ds {{\del \hspace{-6.4pt} \slash}\;}  
\def\ss {{\s \hspace{-6.4pt} \slash}\;}  
\def\ks {{ k \hspace{-6.4pt} \slash}\;}  
\def\ps {{p \hspace{-6.4pt} \slash}\;}   
\def\xs {{x \hspace{-6.4pt} \slash}\;}  
\def\pas {{{p_1} \hspace{-6.4pt} \slash}\;}  
\def\pbs {{{p_2} \hspace{-6.4pt} \slash}\;}   
\def\cFs {{{\cal F} \hspace{-6.4pt} \slash}\;}

\def\Ah{{\hat{A}}}  
\def\Dh{{\hat{D}}}
\def\Gh{{\hat{G}}}
\def\Fh{{\hat{F}}}
\def\Ih{{\hat{I}}} 
\def\Jh{{\hat{J}}} 
\def\Kh{{\hat{K}}}
\def\Lh{{\hat{L}}} 
\def\Ph{{\hat{P}}}
\def\Rh{{\hat{R}}}
\def\Vh{{\hat{V}}} 
\def\Xh{{\hat{X}}}
 
\def\ah{{\hat{\a}}}
\def\bh{{\hat{\b}}}
\def\gh{{\hat{\g}}}
\def\dh{{\hat{\d}}}
\def\hh{\hat{h}}
\def\uh{\hat{u}}  
\def\xh{\hat{x}}  
\def\yh{\hat{y}}  
\def\ph{\hat{p}}  
\def\xih{\hat{\xi}}  
\def\chih{\hat{\chi}}  
\def\Psih{\hat{\Psi}}    
\def\phih{\hat{\phi}}

\def\psit{\tilde{\psi}}  
\def\Psit{\tilde{\Psi}}   
\def\Psibt{\tilde{\bar{Psi}}}  

\def\st{\tilde{\sigma}}  

\def\delt{\tilde{\delta}}
\def\Phit{\tilde{\Phi}}   
\def\Phitb{\overline{\tilde{Phi}}}  
\def\tht{\tilde{\th}}  
\def\lt{\tilde{\l}}
\def\chit{\tilde{\chi}}   
\def\phit{\tilde{\phi}} 

\def\At{\tilde{A}}
\def\Bt{\tilde{B}}
\def\Ct{\tilde{C}}
\def\Dt{\tilde{D}}
\def\Et{\tilde{E}}
\def\Ft{\tilde{F}}
\def\Gt{\tilde{G}}
\def\Ht{\tilde{H}}
\def\It{\tilde{I}}
\def\Jt{\tilde{J}}
\def\Qt{\tilde{Q}}  
\def\Rt{\tilde{R}}  
\def\Mt{\tilde{M }}  
\def\Nt{\tilde{N}}   
\def\St{\tilde{S}}
\def\Vt{\tilde{V}}
\def\Xt{\tilde{X}} 
\def\at{\tilde{a}}
\def\ct{\tilde{c}}
\def\dt{\tilde{d}}
\def\htt{\tilde{h}} 
\def\ft{\tilde{f}}
\def\gt{\tilde{g}}
\def\pt{\tilde{p}}  
\def\qt{\tilde{q}}  
\def\vt{\tilde{v}}  
\def\nt{\tilde{n}}  
\def\ut{\tilde{u}}  
\def\wt{\tilde{w}}  
\def\zt{\tilde{z}} 
\def\xt{\tilde{x}} 
\def\yt{\tilde{y}} 
\def\Psit{\tilde{\Psi}}
\def\vphit{\tilde{\varphi}}  

\def\eb{\bar{\epsilon}} 
\def\delb{\bar{\partial}}  
\def\thb{\bar{\theta}}
\def\mub{\bar{\mu}}
\def\lamb{\bar{\l}}
\def\psib{\bar{\psi}}
\def\sb{\bar{\sigma}}
\def\xib{\bar{\xi}}
\def\chib{\bar{\chi}}

\def\Psib{\bar{\Psi}}
\def\Phib{\bar{\Phi}}
\def\Lamb{\bar{\Lambda}}
\def\Sb{{\overline \Sigma}}
\def\cb{\bar{c}}
\def\hb{\bar{h}}
\def\qb{\bar{q}}
\def\wb{\bar{w}}
\def\ub{\bar{u}}
\def\zb{{\bar{z}}}
\def\Hb{\bar{H}}
\def\Qb{{\bar Q}}
\def\Omegab{\overline{\Omega}}
\def\ob{\overline{\omega}}

\def\Ab{{\overline A}} \def\Bb{{\overline B}} \def\Cb{{\overline C}}  
\def\Db{{\overline D}} \def\Eb{{\overline E}} \def\Fb{{\overline F}}  
\def\Gb{{\overline G}} 
\def\Ib{{\overline I}}  
\def\Jb{{\overline J}} \def\Kb{{\overline K}} \def\Lb{{\overline L}}  
\def\Mb{{\overline M}} \def\Nb{{\overline N}} \def\Ob{{\overline O}}  
\def\Pb{{\overline P}}  \def\Rb{{\overline R}}  
 \def\Tb{{\overline T}} \def\Ub{{\overline U}}  
\def\Vb{{\overline V}} \def\Wb{{\overline W}} \def\Xb{{\overline X}}  
\def\Yb{{\overline Y}} \def\Zb{{\overline Z}}  

\def\fb{{\overline f}}
\def\gb{{\overline g}}
\def\mb{{\overline m}}
\def\lb{{\overline l}}
\def\yb{{\overline y}}
  
\def\ldel{{\overleftarrow{\del}}}
\def\rdel{{\overrightarrow{\del}}}
\def\ldeldel{{\overleftarrow{\del^2}}}
\def\rdeldel{{\overrightarrow{\del^2}}}
\def\ldelb{{\overleftarrow{\bar{\del}}}}
\def\rdelb{{\overrightarrow{\bar{\del}}}}
\def\ba{{\bf a}} 
\def\bk{{\bf k}}  
\def\bl{{\bf l}}  
\def\bp{{\bf p}}  
\def\bq{{\bf q}}  
\def\br{{\bf r}}
\def\bt{{\bf t}}
\def\bu{{\bf u}}
\def\bv{{\bf v}}
\def\bx{{\bf x}}  
\def\by{{\bf y}}  
\def\bR{{\bf R}}  
\def\bV{{\bf V}}

\def\bone{{\bf 1}}  

\def\va{{\vec a}}
\def\vk{{\vec k}}
\def\vp{{\vec p}}
\def\vq{{\vec q}}
\def\vx{{\vec x}}
\def\vy{{\vec y}}
\def\vu{{\vec u}}
\def\vv{{\vec v}}
\def \vH{{\vec H}}
\def \vg{{\vec g}}

\def\vs{{\vec \sigma}}
\def\vtau{{\vec \tau}}

\newcommand{\ov}[1]{\overrightarrow{#1}}

\def\frA{\mathfrak{A}}
\def\frB{\mathfrak{B}}
\def\frC{\mathfrak{C}}
\def\frD{\mathfrak{D}}
\def\frE{\mathfrak{E}}
\def\frF{\mathfrak{F}}
\def\frG{\mathfrak{G}}
\def\frH{\mathfrak{H}}
\def\frM{\mathfrak{M}}
\def\frN{\mathfrak{N}}
\def\frR{\mathfrak{R}}
\def\frW{\mathfrak{W}}

\def\fra{\mathfrak{a}}
\def\frb{\mathfrak{b}}
\def\frf{\mathfrak{f}}
\def\frg{\mathfrak{g}}
\def\frh{\mathfrak{h}}
\def\frl{\mathfrak{l}}
\def\frs{\mathfrak{s}}
\def\fri{\mathfrak{i}}
\def\frj{\mathfrak{j}}

\def\ma{\mathfrak{a}}
\def\mg{\mathfrak{g}}
\def\mh{\mathfrak{h}}
\def\mR{\mathfrak{R}}
\def\mN{\mathfrak{N}}

\def\d{\delta}\def\D{\Delta}\def\ddt{\dot\delta}  
  
\def\pa{\partial} \def\del{\partial}  
\def\xx{\times}  
\def\uno{\mbox{1 \kern-.59em {\rm l}}}    
  
\def\trp{^{\top}}  
\def\inv{^{-1}}  
\def\dag{{^{\dagger}}}  
\def\pr{^{\prime}}  
  
\def\rar{\rightarrow}  
\def\lar{\leftarrow}  
\def\lrar{\leftrightarrow}  
  
\newcommand{\0}{\,\!}      %this is just NOTHING!  
\def\one{1\!\!1\,\,}  
\def\im{\imath}  
\def\jm{\jmath}  
  
\newcommand{\tr}{\mbox{tr}}  
\newcommand{\slsh}[1]{/ \!\!\!\! #1}  
  
\def\vac{|0\rangle}  
\def\lvac{\langle 0|}  
  
\def\hlf{\frac{1}{2}}  
\def\ove#1{\frac{1}{#1}}  

\def\Box{\square}  
\def\CC {\mathbb{C}}
\def\FF {\mathbb{F}}
\def\RR{\mathbb{R}}
\def\NN{\mathbb{N}}  
\def\ZZ{\mathbb{Z}}  
\def\bb#1{{\bf #1}}  
\def\bcomment#1{}  
\def\bfhat#1{{\bf \hat{#1}}}  
\def\VEV#1{\left\langle #1\right\rangle}  

\newcommand{\ex}[1]{{\rm e}^{#1}} \def\ii{{\rm i}}  

\newcommand{\lrbrk}[1]{\left(#1\right)}
\newcommand{\lrsbrk}[1]{\left[#1\right]}
\newcommand{\sfrac}[2]{{\textstyle\frac{#1}{#2}}}
 
\def\stw{{\sqrt{2}}}

\def\rf {{\rm f}}
\def\ri {{\rm i}}
\def\rj {{\rm j}}
\def\rk {{\rm k}}
\def\rl {{\rm l}}
\def\rs {{\scriptscriptstyle \rm S}}
\def\rt {{\scriptscriptstyle \rm T}}

\def\rQ {{\scriptscriptstyle \rm \cQ}}
\def\rR {{\scriptscriptstyle \rm \cR}}

\def\cQb{{\cal \Qb}}
\def\cRb{{\cal \Rb}}
\def\cWb{{\cal \Wb}}

\def\fd {{\rm N}}
\def\afd {{\overline{\rm N}}}

\def \II {I\hspace{-.1em}I\hspace{.1em}}
\def \IIA {\mbox{\II A\hspace{.2em}}}
\def \IIB {\mbox{\II B\hspace{.2em}}}
\def \gs {g^s}
\def \ls {\lambda^s}

\def \I {{\cal I}}
\def \qs {q\hspace{-.53em}/\hspace{.15em}}
\def \ks {k\hspace{-.53em}/\hspace{.15em}}
\def \YM {{\mbox{\tiny YM}}}
\def \gym {g_{\YM}}

\def \Lc {\L_c}
\def\IR{\relax{\rm I\kern-.18em R}}
\def \id {{\bf 1}}

\def\cci{\ell}
\def\ccj{\ell'}

\def \thbb{\overline{\th\th}}
\newcommand \ol{\overline}
\def \lamb{\bar{\lambda}}
\def \vphi{\varphi}
\def \lambh{\hat{\bar{\lambda}}}
\def \lh{\hat{\lambda}}
\def \dd{\ddagger}

\newcommand{\QNB}[3]{[#1,#2,#3]}
\def\hm{\tilde{\eta}} %%"eta tilde", aka "eta minus"
\def\lp{l_{+}}
\def\lm{l_{-}}
\def \PS {{(\text{PS})}}
\def \Dir {{(\text{Dirac})}}
\def \WY {{(\text{WY})}}
\def \Sin {{(\text{Sin})}}
\def \tHP{{(\text{'t-P})}}
\def \uo {{U(1)}}
\def \Lt {\tilde{L}}
\def \tn {{\tau}^{(n)}}
\def \rn {{\hat{r}}^{(n)}}
\def \thn {{\hat{\theta}}^{(n)}}
\def \vphin {{\hat{\vphi}}^{(n)}}

\newcommand{\Ga}{{\Gamma}}
\newcommand{\De}{{\Delta}}
\newcommand{\Lm}{{\Lambda}}
\newcommand{\Om}{{\Omega}}

\newcommand{\al}{{\alpha}}
\newcommand{\ga}{{\gamma}}
\newcommand{\de}{{\delta}}
\newcommand{\ep}{{\epsilon}}
\newcommand{\vep}{{\varepsilon}}
\newcommand{\te}{{\theta}}
\newcommand{\ka}{{\kappa}}
\newcommand{\vpi}{{\varpi}}
\newcommand{\sig}{{\sigma}}
\newcommand{\om}{{\omega}}

\newcommand{\alt}{{\rm alt}}
\newcommand{\bdy}{{\rm bdy}}
\newcommand{\bsa}{{\boldsymbol{a}}}
\newcommand{\bsb}{{\boldsymbol{b}}}
\newcommand{\bsD}{{\boldsymbol{D}}}
\newcommand{\bsk}{{\boldsymbol{k}}}
\newcommand{\bsM}{{\boldsymbol{M}}}
\newcommand{\bulk}{{\rm bulk}}
\newcommand{\cont}{{\rm cont.}}
\newcommand{\cdN}{{\mathcal{N}_d}}
\newcommand{\lan}{{\langle}}
\newcommand{\pd}{{\partial}}
\newcommand{\R}{{\rm R}}
\newcommand{\rad}{{\rm rad}}
\newcommand{\ran}{{\rangle}}
\newcommand{\Slash}[1]{{\ooalign{\hfil/\hfil\crcr$#1$}}} 
\newcommand{\srel}[2]{{\stackrel{\scriptstyle #1}{\scriptstyle #2}}}
\newcommand{\std}{{\rm std}}
\newcommand{\U}{{\rm U}} 
\newcommand{\ul}{\underline}
\newcommand{\UV}{{\rm UV}}
\newcommand{\wg}{{\wedge}}
\newcommand{\wh}{\widehat}

\def\Log{\mathop{\rm Log}}
\def\Spin{\mathop{\rm Spin}}
\def\SO{\mathop{\rm SO}}
\def\O{\mathop{\rm O}}
\def\SU{\mathop{\rm SU}}
\def\U{\mathop{\rm U}}
\def\Sp{\mathop{\rm Sp}}
\def\SL{\mathop{\rm SL}}
\def\GL{\mathop{\rm GL}}

\def\det{\mathop{\rm det}\nolimits}
\def\sign{\mathop{\rm sign}\nolimits}
\def\mod{\mathop{\rm mod}\nolimits}
\def\tr{\mathop{\rm tr}\nolimits}
\def\diag{\mathop{\rm diag}\nolimits}
\def\Re{\mathop{\rm Re}\nolimits}
\def\Im{\mathop{\rm Im}\nolimits}
\def\Tr{\mathop{\rm Tr}\nolimits}
\def\bbra{{\langle\kern-2.5pt\langle}}
\def\kket{{\rangle\kern-2.5pt\rangle}}
\def\Bbra{{\Big\langle\kern-3.5pt\Big\langle}}
\def\Kket{{\Big\rangle\kern-3.5pt\Big\rangle}}

\author{Chong-Sun Chu$^{1,2,3}$ 
and Hiroshi Isono$^1$ \\
$^1$ Department of Physics,
National Tsing-Hua University, \\
Hsinchu 30013, Taiwan \\
$^2$ National Center for Theoretical Sciences,
National Tsing-Hua University, \\
Hsinchu 30013, Taiwan \\
$^3$ Centre for Particle Theory and Department of Mathematics, \\
Durham University, Durham, DH1 3LE, UK\\
E-mail:  
\email{chong-sun.chu@durham.ac.uk},
\email{hisono@phys.nthu.edu.tw}
}

\title{
Instanton String and M-Wave in Multiple M5-Branes System
}

\abstract{
We consider the non-abelian self-dual two-form theory 
\href{http://arxiv.org/abs/arXiv:1203.4224}{\cite{CK}}
and find 
new exact  solutions. 
Our solutions are supported by Yang-Mills (anti)instantons
in 4-dimensions and describe wave moving in null directions. 
We argue and 
provide evidence that these instanton string solutions 
correspond to M-wave (MW) on the worldvolume of
multiple M5-branes. When dimensionally reduced on a circle, 
the MW/M5 system is reduced to the D0/D4 system with the D0-branes 
represented by  
the Yang-Mills instanton of the D4-branes Yang-Mills gauge theory. 
We show that this picture is precisely reproduced by 
the dimensional reduction of our instanton string solutions.
}

\preprint{DCPT-13/21}

\keywords{M-Theory, D-branes, M-branes, Gauge Symmetry}

\begin{document}

\section{Introduction}

%c5 The low energy 
The theory of $N$ coincident M5-branes in a flat spacetime 
is given by an interacting (2,0) superconformal
theory in six dimensions \cite{witten0}. The understanding of the
dynamics of this system is of utmost importance.
On general grounds, the theory does not have a free 
dimensionless parameter and  is inherently non-perturbative. 
It does not mean
that an action does not exist, 
%c5
though it does mean that the use of the action
will be limited to non-perturbative analysis, for example, the studies of
solutions to the equations of motion. This is still very interesting, 
particularly since
much of the spacetime M-theory physics
can be learnt from the physics of the solitonic objects of the 
the worldvolume theory of M5-branes.
This is of course 
valid also for M2-branes as well as 
D-branes in string theory and this kind of target space-worldvolume
duality has dominated the developments of string theory in the last 15 years 
or so, with the AdS/CFT correspondence \cite{adscft} 
being the most 
%c5
celebrated duality.

For a single M5-brane, the equation of motion
has been constructed \cite{hs,PS,schw1,pst}.
For multiple M5-branes, a major difficulty has been 
to non-abelianize the self-dual
tensor gauge dynamics. This problem was tackled and 
%c5 solved 
a solution was presented in \cite{CK}
where a consistent self-duality equation of motion for the non-abelian tensor 
gauge field was constructed. Moreover it was proposed as the
%c5 
low energy 
equation of motion of the self-dual tensor field 
living on the worldvolume of a system of multiple M5-branes.

The construction of \cite{CK} consists of two 
primary steps: the
identification of the
form of the
non-abelian gauge symmetry and the construction of the
dynamical self-duality equation. The 
%c5
proposed form of the
non-abelian gauge symmetry was  motivated by the
analysis in \cite{CS} where
a set of 5-dimensional non-abelian one-form gauge fields
was introduced in order to
incorporate non-trivial interactions among the 2-form potentials \cite{chu}.
The introduction of the non-abelian 1-form
gauge fields allows one to write covariant derivative and
to introduce non-abelian transformation
for the fields
\footnote{
Similar forms of gauge symmetry were also considered by 
%h3 three-form -> 3-form
\cite{ho}, as well as \cite{sezgin1} where an extra 3-form potential 
%h3 one-form -> 1-form
was introduced in addition to the propagating 1-form gauge potential.
The latter formulation were further developed in \cite{sezgin2,d1} 
and provides a construction for 
a class of (1,0) superconformal models in 6-dimensions.}. 
However since there is no room for propagating
1-form gauge fields in the (2,0) self-dual
tensor multiplet in 6-dimensions, they must be constrained and be
non-propagating
\footnote{
The philosophy is very similar to the BLG \cite{BLG}
and ABJM model \cite{ABJM} of multiple M2-branes where a set of
non-propagating Chern-Simons gauge fields was introduced in order to
allow for a simple representation of
the highly non-linear and non-local self interactions of the matter fields
of the $\cN=8$ supermultiplet in 3-dimensions.}.

The constraint needed for the specification of the gauge symmetry was 
identified in \cite{CK}. There, a self-duality equation 
for a non-abelian 2-form in 6-dimensions 
was constructed in the gauge $B_{5\m} =0$ ($\m =0,\cdots, 4$) and 
reads
\be\label{sd-na}
\Ht_{\m\n} = \pa_5 B_{\m\n}.
\ee
The constraint for the gauge field $A_\m$ is given by
\be \label{FH}
F_{\m\n} = c\int dx_5 \Ht_{\m\n}.
\ee
Here
\be
H_{\m\n\r}= D_{[\m}B_{\n\r]} = \pa_{[\m}B_{\n\r]}+[A_{[\m},B_{\n\r]}],
\ee
\be
\tilde{H}_{\m\n} = -\ove{6}\e_{\m\n\r\s\t}H^{\r\s\t},\qquad \e_{01234}=-1,
\ee
\be
F_{\m\n} = \pa_{\m} A_{\n} - \pa_{\n} A_{\m} + [A_\m,A_\n]
\ee
%c5
and $c$ is a constant that is fixed by quantization condition of the 
self-dual strings solution of the theory \cite{CKV,CV}.
All fields are in the adjoint representation of the Lie algebra of 
the gauge group $G$. 
The self-duality equation \eq{sd-na} and the constraint \eq{FH}
were derived in \cite{CK} as the equations of motion of an action
principle, which is a generalization of the abelian theory of
Perry-Schwarz \cite{PS}. Evidence that this self-duality equation describes
%h3 were -> was
physics of M5-branes was provided in \cite{CK}, and further in 
\cite{CKV,CV} where non-abelian self-dual string solutions were constructed.
In these latter two papers,  
it was argued that  the form of the 1/2 BPS equation with a single scalar 
activated could be derived from the requirement of conformal symmetry and 
R-symmetry of the system and it was shown that the solution of
the pure gauge sector
could be lifted to become  a solution of the non-abelian (2,0) theory
with  self-dual electric and magnetic charges.
In M-theory, the 
self-dual string arises from the intersection of a system of 
M2-branes with the system of multiple M5-branes. It is 
satisfying that the constant $c$ is fixed, which would otherwise be 
a free dimensionless constant in the theory and hence contradicts with
what we know about M5-branes in flat space.
%c5 , by the charge quantization of the  self-dual string solution. 
It is also
quite encouraging that a complete agreement \cite{CV} of the 
field theory results and the supergravity descriptions \cite{siampos} 
was found, thereby providing substantial support to the proposal
of \cite{CK} that 
\eq{sd-na} is the self-duality equation for multiple M5-brane system.

We remark that 
in these solutions, the self-dual strings were uniform and static and lie 
in the $x^0$ and $x^4$ directions. As a result, 
the solutions take the form with $B = B(x^i, x^5)$,
$ A= A(x^i)$,  $i =1,2,3$. It is interesting that 
the self-dual string solutions found in \cite{CKV, CV} were supported by having its 
auxiliary gauge field $A$ given by a 
non-abelian monopole configuration in 3-dimensions $x^i$, 
%h3 and the -> and that the
and that the charge of the
self-dual strings is given by the monopole charge. Motivated by this observation,
it is natural to ask if there are other exact solutions of the non-abelian 
self-duality equation \eq{sd-na} that are supported by other interesting 
gauge field  configurations; and if so, what are their 
interpretations in M-theory? 

In the next section, we construct a 
%c5
new class of solutions of the 
self-duality equation 
%c5
obtained by having its auxiliary gauge fields given by 
Yang-Mills (anti-)instantons 
in 4-dimensions. We will argue that the solutions describe M-wave on M5-branes.
The paper is concluded with some further discussions in section 3. 

\section{M-Wave Solution}

Throughout the paper we follow the convention of \cite{CK,CKV,CV}. 
In particular the 5d and 6d coordinates are denoted 
by $x^{\mu}=(x^0,x^1,x^2,x^3,x^4)$ and $x^M=(x^{\mu},x^5)$.
We are interested in finding new solutions of the self-duality equation
and 
%c5 its
their M-theory interpretation.
Let us first consider the case of non-compact $x^{5}$. The equations read
%h3 commas and full stops are added
\be
  \Ht_{\mu\nu} = \pd_5 B_{\mu\nu}, \label{eom1} 
\ee
\be
  F_{\mu\nu} = c[B_{\mu\nu}(x_5=\infty)-B_{\mu\nu}(x_5=-\infty)]. \label{eom2} 
\ee
Splitting $x^{\mu}$ into $x^0$ and $x^a=(x^1,x^2,x^3,x^4)$, 
then  \eqref{eom1} reads
%h3 commas and full stops are added
\begin{align}
  \Ht_{0a} &= \pd_5B_{0a} = -\frac{1}{6}\ep_{abcd}H_{bcd}, \label{eom3} \\
  \Ht_{ab} &= \pd_5B_{ab} = \frac{1}{2}\ep_{abcd}H_{0cd}. \label{eom4}
\end{align}
Let us take an ansatz $B_{0a}=0, A_0=0$. We have
\begin{align}
  0 &= -\frac{1}{6}\ep_{abcd}H_{bcd}, \label{eom5} \\
  \pd_5B_{ab} &= \frac{1}{2}\ep_{abcd}\pd_0B_{cd}. \label{eom6}
\end{align}
The equation \eqref{eom5} can be solved with 
\begin{align} 
  B_{ab} = F_{ab}\; f(x^0,x^5) \label{sol}
\end{align}
for arbitrary $A_a$ due to the Bianchi identity of $F_{ab}$. If we further 
assume that $A_a$'s are independent of $x^0$, then
\eqref{eom6} reads
\be
  F_{ab}\pd_5f = \frac{1}{2}\ep_{abcd}F_{cd}\pd_0f.
\ee
This can be solved with
\begin{align}
  F_{ab} \mbox{ being SD and } &f=f(x^0+x^5), \label{c1}\\
  \mbox{or, }  F_{ab} \mbox{ being ASD and } & f=f(x^0-x^5), \label{c2}
\end{align}
where (A)SD stands for (anti)self-dual. 
Finally the equation
\eqref{eom2} requires that
\be
c [f(\infty)-f(-\infty)] = \pm 1,
\ee
where $+$ is for the SD case and $-$ is for the ASD case.
For example, one can take $f =  \tanh (x^0 \pm x^5) /(2c)$.  
Since our solution is supported by an instanton configuration, it might 
be called an instanton string. What kind of M-theory object does it describe?

In M-theory, there are four basic objects preserving half of the 
supersymmetries.
These are the M5-brane, M2-brane, M-wave (MW) and the 
Kaluza-Klein monopole (MK). 
The M-wave solution in supergravity was originally constructed in \cite{hull}.
For objects with an extended longitudinal space, 
it is possible to turn on a wave in a null direction along the brane. 
For example, 
one can have an MW in the $x^0, x^5$ directions on a system of M5-branes. The 
intersecting brane system is 1/4 BPS in the 11-dimensional supergravity and a
smeared 
%c4 (localized) 
(delocalized) supergravity solution has been constructed in \cite{tseytlin}. 
We propose that our instanton string solution is describing a MW on a system
of multiple M5-branes.

To check this, let us consider supersymmetry. 
We remark that the abelian $(2,0)$ 
tensor multiplet has the supersymmetry transformation
\begin{align}
  \de X^I &= i\overline{\ep}\Ga^I\Psi, \label{sudsy1} \\
  \de \Psi &= \Ga^M\Ga_I\pd_MX^I\ep+\frac{1}{3!}
\frac{1}{2}\Ga^{MNL}H_{MNL} \ep,   
   \label{susy2} \\
  \de B_{MN} &= i\overline{\ep}\Ga_{MN}\Psi, \label{susy3} 
\end{align}
where $I,J=6,7,\cdots,10$ and $M,N,L=0,1,\cdots,5$. For the 
non-abelian case, let us assume that the supersymmetry transformation
takes on a similar form plus terms that vanish on our solution. 
Specifically,
%c5 
let us 
assume that for $\Psi$,
\be
  \de \Psi = \Ga^M\Ga_ID_MX^I\ep+
\frac{1}{3!}\frac{1}{2}\Ga^{MNL}H_{MNL} \ep + \cdots,  \label{susy2p} 
\ee
where $\cdots$ denotes terms that vanish for our solution, e.g. terms involving
two or more different $X$'s. This form of the supersymmetry has been tested in 
\cite{CKV,CV}.
The form \eq{susy3} of supersymmetry is also reasonable
since, as a result of our ansatz \eq{sol}, it implies $\de F_{ab} = 
\de B_{ab}=0$ for our solution and this is consistent with the fact that
the auxiliary gauge field $A_a$ is supposed to be supersymmetry invariant.

Now, for our solution $X^I=0, \Psi=0$ hold, 
therefore \eqref{susy2p} implies
\begin{align}
  0 = (\Ga^{0ab}H_{0ab}+\Ga^{5ab}H_{5ab})\ep, \label{susy4}
\end{align}
where $H_{5ab}:=\pd_5B_{ab}$.
Since our solution satisfies $H_{0ab}=\pm H_{5ab}$
with the plus sign for SD $F_{ab}$ and the minus sign for ASD $F_{ab}$,
therefore \eqref{susy4} implies
\begin{align}
  \Ga_{05}\ep = \pm\ep \label{susy5}
\end{align}
with the plus sign for the SD case and the minus sign for the ASD case.
The projector condition \eqref{susy5} means a breaking of  
half of the supersymmetry of the $(2,0)$ theory and is in fact precisely 
the same
supersymmetry preserving condition for 
%c5 the MW in supergravity.
MW on M5-branes. 
This supports the
identification of our instanton string with the MW. 

To  further justify our claim, let us consider a compactification of 
our system on 
a circle, say, for  $x^5$ to be compactified on a circle of radius $R$.
In the brane picture, the MW/M5 system is reduced to a D0/D4 system where
the D0-branes are represented as Yang-Mills instanton of 
the Yang-Mills theory of the D4-branes
\cite{douglas,witten} due to the coupling on the D4-brane worldvolume
to the RR-gauge field $C_1$,
\be \label{RR}
S_{RR} \sim \int d^5 x \;  C_1 \wedge \tr(F \wedge F).
\ee
The number of D0-branes is given by the instanton number. 
We will now show that this picture is reproduced
precisely by the compactification of our instanton string solution.

For $x^5$ being compactified on a circle of radius $R$,
the condition \eqref{eom2} reads
%h3 comma added
\begin{align} 
  F_{\m\n} = 2\pi Rc H^{(0)}_{5\m\n}, \label{eom2c}
\end{align}
where $H^{(0)}_{5\m\n}$ is the zero mode part 
of the modes expansion of $H_{5\m\n}$. 
For our ansatz with nonzero 
$A_a$ and $B_{ab}$, the equations \eqref{eom5}, \eqref{eom6} read
\bea
  & \ep_{abcd}(D_b B_{cd})^{(n)} = 0, \label{eom7} \\
  & (\pd_5B_{ab})^{(n)} = \frac{1}{2}\e_{abcd}(\pd_0B_{cd})^{(n)}, \label{eom8}
\eea
where $X^{(n)}$ denotes coefficient of the $n$-th 
Fourier mode of 
$X = \sum_{n= -\infty}^\infty X^{(n)} e^{inx^5/R}$. 
Consider the ansatz 
\begin{align}
B_{ab} = \a\times 
(x^5 \pm x^0) F_{ab}(x^a) + \sum_{n=-\infty}^{\infty} 
B^{(n)}_{ab}(x^0,x^a)e^{inx^5/R},
\end{align}
where $\al$ is a constant,
$F_{ab}$ is SD for the plus sign and ASD for the minus sign.
Note that a linear term in $x^0$ and $x^5$ is allowed as 
it is only required that 
the field strength to be a periodic function of $x^5$. 
%c5
The form of the linear part is fixed by solving the zeroth part of the 
equations \eq{eom7}, \eq{eom8}.
%c5
As for the Fourier modes, the zeroth mode is required 
by  \eq{eom7} to obey
$ \ep_{abcd} D_b  B_{cd}^{(0)} =0$ and this can be 
solved by 
\be
B_{ab}^{(0)} = \b_0 F_{ab},
\ee
where $\b_0$ is a constant. 
%c5 This certainly solves the  zero mode part of the  equation \eq{eom8}. 
As for the non-zero mode equations, they can be solved with
\be
B_{ab}^{(n)} = \b_n  e^{\pm inx^0/R} F_{ab},
\ee
where $F_{ab}$ is SD for the plus sign and ASD for the minus sign
and $\b_n$ are constants.

%c5 
All in all, the dimensionally reduced system \eqref{eom7}, \eqref{eom8}
can be solved by
\be
B_{ab} = F_{ab}(x^k)f(x^0 \pm x^5),
\ee
%c5 where
with 
\be
f(x^0 \pm x^5) := \a (x^5 \pm x^0)+\sum_{n=-\infty}^\infty \b_n   
e^{\frac{in (x^5 \pm x^0 )}{R}}, \label{solc}
\ee
where $F_{ab}$ is SD for the plus sign and ASD for the minus sign.
This is of course nothing but simply the solution 
\eq{sol}, \eq{c1}, \eq{c2} with now a profile function 
$f$ that is consistent with the 
compactification condition. The constant $\a$ can be fixed by noticing that
our solution \eq{solc}  gives 
\be
H^{(0)}_{5ab}=\a F_{ab}
\ee 
and the condition \eqref{eom2c} 
%c5 thus 
implies that
\begin{align}
\a = \frac{1}{2\pi Rc}.
\end{align}
Note that $H^{(0)}_{5ab}$ is independent of $B^{(0)}$. 
We also record
\be \label{H-KK}
H_{5ab} = F_{ab} \; \big(\a +\sum_{n=-\infty}^\infty 
\frac{i n \b_n}{R}
e^{\frac{in (x^5 \pm x^0 )}{R}} \big).
\ee

%c5 
Now the general analysis of the dimensional reduction of the 
non-abelian self-duality equation has been performed in \cite{CK}.
Note that on dimensional reduction of our non-abelian M5 brane theory, 
the field strength 
\be
 F_{\mu\nu} = 2\pi Rc H^{(0)}_{5\mu\nu}
\ee 
becomes the field strength of 
%c5 the Yang-Mills $A_{\mu}$ of 
the 5d supersymmetric Yang-Mills
theory.
Indeed it is no longer interpreted
as an auxiliary field but becomes propagating and 
obeys a dynamical equation of motion \cite{CK},
\begin{align}
 D_{\mu}F^{\mu\nu} &= -\frac{\pi R}{2}\e^{\nu\a\b\g\d}[F_{\a\b}, B_{\g\d}].
\label{eom5d}
\end{align} 
In general, this is more complicated than the source free Yang-Mills equation. 
In
\cite{CK}, the additional term on the right hand side 
was argued to be arising from higher derivatives corrections of the Yang-Mills 
action. For our specific solution \eq{solc}, 
the 
right hand side of \eqref{eom5d} is zero since $B_{ab} \propto F_{ab}$
in our solution.  Therefore, on dimensional reduction, our
instanton string gives rise 
%c4 precisely immediately 
%c5 
exactly to an instanton (or anti-instanton) 
configuration in a standard 5-dimensional Yang-Mills action.
This is precisely
what is expected for the system of D0-branes in D4-branes.
Besides, we can see that the RR coupling \eq{RR} does arise naturally from 
the non-abelian M5-branes theory on dimensional reduction. 
In fact, \eq{RR} is just a part of the coupling of the metric 
to the energy-momentum tensor
\be
\int d^6 x \;  g_{05} T^{05}
\ee
since $C_1 \sim  g_{05}$ and quite
generally one can expect the energy-momentum tensor of the system of multiple 
M5-branes to contain the term 
%c5 
(up to a normalization constant), 
\be \label{T}
T^{MN} = \tr (H^{MPQ} H^N{}_{PQ}) + \cdots.
\ee 
This is the case for the abelian M5-brane \cite{t1} and
for the non-abelian (2,0) theory \cite{t2}. 
%c4 
And it should be true irrespective
of whether the theory is self-dual or not.

\section{Discussions}

In this paper, we have constructed a new solution of the non-abelian 
self-duality 
equation. The solution is supported by an instanton or anti-instanton 
configuration for the auxiliary Yang-Mills gauge field. 
We have argued and provided 
evidence that our solution provides 
a field theory description of the M5/MW intersecting
system. Together with other evidence provided previously \cite{CKV,CK,CV},
we are quite confident of \eq{sd-na}, \eq{FH}
as providing a description of the self-dual dynamics of multiple M5-branes. 

As instanton, our solution is localized in the Euclidean $x^a$ directions. 
It will be interesting to construct localized supergravity 
solution for the M5/MW  system. 
It may be possible to build it similarly to 
the localized supergravity solution of 
intersecting M5-branes \cite{loc-M5}. 
This will provide further features of the 
%c5 system 
solution which one can compare with 
the field theory results,
%c5  and one can perform 
thus furnishing a more detailed and more interesting 
test.

It is important to complete the bosonic self-duality equations 
with supersymmetry. Some difficulties associated with the
supersymmetric completion have been discussed in \cite{CK}. 
In our present construction of the instanton string solution,
we have provided another valuable piece of information concerning the
form of the supersymmetry transformation. It will be important to push
this direction forward. 

If a $C$-field is turned on in the worldvolume of the system of M5-branes, 
then one 
can expect that 
%c5 a 
some kinds of star-product will arise as a result of the emerged 
noncommutative geometry, and 
the self-duality equation will be modified. In \cite{qg}, 
the logic was reversely
applied and it was shown how one can deduce the known form of 
worldvolume noncommutative geometry of D-branes 
\cite{ncg1,ncg2,ncg3} 
from a knowledge of how the $B$-field modifies the Nahm equation; 
as well as the
noncommutative geometry of M5-brane in a large $C$-field from a 
knowledge of how the $C$-field modifies the Basu-Harvey equation \cite{BH}. 
The found noncommutative geometry takes the form of a quantum Nambu geometry 
\cite{CG}. However there has not been success in reformulating this 
operatorial geometry in terms of more familiar language of a deformed
star-product.
Knowing  how the $C$-field modifies the nonabelian 
self-duality equation may provide hints to this problem.

In addition to the 5d-SYM proposal \cite{dou,lam2},
there exist a number of other earlier proposals \cite{qm1,qm2} and \cite{dec}
for the  definition of the  theory of multiple M5-branes.
These proposals have the advantage of being supposed to be providing a fundamental
quantum description, but is however much less explicit.
It will be interesting to explore the connection of our description with 
these proposals.

In some sense, 
the self-duality equation \eq{sd-na} is a higher dimensional 
generalization of the 
%c5
self-dual Yang-Mills equation in 4-dimensions.
The self-dual Yang-Mills equation is an interesting  
%c4
mathematical physics system.
Apart from  numerous important applications in mathematics, it is 
exactly solvable. Moreover, it has been
conjectured by Ward \cite{ward}
to be a kind of master equation, which states that
all the integrable equations in three of lower dimensions can be obtained from the 
self-dual Yang-Mills equation by a reduction.
It is interesting to 
ask if the  self-duality equation \eq{sd-na} has also similarly rich and interesting
mathematical properties. 
For example, is it integrable? is there a twistor construction behind, and is
there any connection with the twistor or loop formulation proposed in 
\cite{twistor}? 
We leave these interesting questions
for further exploration.

\section*{Acknowledgements}

It is our pleasure to thank Kazuyuki Furuuchi, 
Sheng-Lan Ko, Christian Saemann, Richard Szabo,  
Pichet Vanichchapongjaroen and Martin Wolf
for discussions. CSC is supported in part 
by the STFC Consolidated Grant ST/J000426/1 and 
by the grant 
101-2112-M-007-021-MY3
of the National Science Council, Taiwan.

\end{document}